\input phyzzx
\nopagenumbers

%
\def\refout{\par\penalty-400\vskip\chapterskip
   \spacecheck\referenceminspace
   \ifreferenceopen \Closeout\referencewrite \referenceopenfalse \fi
   \noindent{\bf References\hfil}\vskip\headskip
   \input \jobname.refs }

\def\IR{\relax{\rm I\kern-.18em R}}

\def\npb#1#2#3{{\it Nucl. Phys.} {\bf B#1} (#2) #3 }
\def\plb#1#2#3{{\it Phys. Lett.} {\bf B#1} (#2) #3 }

\def\mpla#1#2#3{{\it Mod. Phys. Lett.} {\bf A#1} (#2) #3 }

\def\jmp#1#2#3{{\it J. Math. Phys.} {\bf #1} (#2) #3 }

\def\jpa#1#2#3{{\it J. Phys.} {\bf A#1} (#2) #3}

\REF\Klebreview{I. R. Klebanov, {\it ``String theory in two
dimensions'',} in ``String Theory and Quantum Gravity'',
Proceedings of the Trieste Spring School 1991, eds. J. Harvey et al.,
(World Scientific, Singapore, 1992).}
\REF\Kutreview {D. Kutasov, {\it ``Some properties of (non) critical
strings'',} in ``String Theory and Quantum Gravity'',  Proceedings of
the Trieste Spring School 1991, eds. J. Harvey et al., (World Scientific,
Singapore, 1992).}
\REF\Polone {J. Polchinski, \npb {346}{1990}{253.}}
\REF\DaJe {S.~R. Das and A. Jevicki, \mpla {5} {1990} {1639.}}
\REF\AvJe {J. Avan and A. Jevicki, \plb {266}{1991}{35;}
\plb {272}{1991}{17.}}
\REF\winf {D. Minic, J. Polchinski and Z. Yang, \npb {369}{1992}{324;}}
\REF\JeYo {A. Jevicki and T. Yoneya, \npb {411}{1994}{64.}}
\REF\KlebPol {I.~R. Klebanov and A.~M. Polyakov, \mpla {6}{1991}{3273.}}
\REF\JRvT {A. Jevicki, J.~P. Rodrigues and A. van Tonder,
\npb {404} {1993} {91.}}
\REF\Ginsparg{P. Ginsparg, {Les Houches Summer School 1988, 1-168}.}
\REF\Zha{C.-H. Zha, \jmp {35}{1994}{35.}}
\REF\BaFlo{L. Baulieu and E.G. Floratos, \plb {258}{1991}{171.}}
\REF\McFBP{{A. Macfarlane, \jpa {22}{1989}{4581; }}
{L. Biedenharn, \jpa {22}{1989}{L873; }}
{A. Polychronakos, \mpla {5}{1989}{2325.}}}
\REF\Witten {E. Witten, \npb {373}{1992}{187.}}
\REF\ChaiDem{M. Chaichian and A.P. Demichev, {\it ``Path integrals
with generalized grassmann variables",} Hu-SEFT-R-1995-09.}
\REF\BabGerAGS{{O. Babelon, \plb {215}{1988}{523; }}
{J.L. Gervais, \npb {391}{1993}{287; }}
{L. Alvarez-Gaum\'e, C. G\'omez and G. Sierra, \npb {319}{1989}{155.}}}
\REF\OLough{M. O'Loughlin, \npb {319}{1989}{155.}}

\singlespace
\hsize=6.0in
\vsize=8.5in
\voffset=0.0in
\hoffset=0.0in
\overfullrule=0pt

\line{}
\line{\hfill BROWN-HET-1031}
\line{\hfill January 1996}
\vskip .75in
\centerline{{\bf FINITE [Q-OSCILLATOR] REPRESENTATION}}
\centerline{{\bf OF 2-D STRING THEORY}}
\vskip .40in
\centerline{ANTAL JEVICKI and ANDR\'e VAN TONDER
 \foot{Work started while supported by a German Science Foundation
(DFG) postdoctoral grant and the Institute of Physics of the University
of Hannover, Germany.  Further support provided through
a South African FRD postdoctoral grant.}}
\smallskip
\centerline{{\it Physics Department, Brown University}}
\centerline{{\it Providence, Rhode Island 02912, USA}}
\vskip .25in
\centerline{\bf Abstract}

\noindent
We present a simple physical representation for states
of two-dimensional string theory.  In order to incorporate a fundamental cutoff
of the order $1/g_{\rm st}$ we use a picture consisting of q-oscillators at the
first
quantized level. In this framework we also find a representation
 for the (singular)
negatively dressed states representing nontrivial string backgrounds.

\vskip 0.70in

{\chapter {Introduction}}
\medskip

Two-dimensional string theory [\Klebreview,\Kutreview] possesses a most
interesting structure.
Its spectrum contains, in addition to a massless scalar particle (the tachyon)
[\Polone,\DaJe], also an infinite sequence of discrete states
with a closely related $W_\infty$ symmetry algebra [\AvJe,\winf].
In the world sheet
picture with $X_\mu = (X, \varphi)$, this spectrum is given by the vertex
operators
$$
  \Psi^{(\pm)}_{Jm} \equiv
    \left( H_-(z)\right)^{J-m}:e^{iJX(z)}:e^{(-1\pm J)\varphi(z)}:
    \eqn \voprep
$$
with discrete momenta $p_X = m$, $p_\varphi = \pm J$.

Of some importance is the difference between the states of positive
Liouville dressing $\Psi^{(+)}$ versus those of negative dressing
$\Psi^{(-)}$.
The former are physical states.  In particular, for tachyons they
provide
left ($p_X > 0$) and right ($p_X < 0$) moving scattering states.
The negatively dressed states  to which we shall refer as ``singular"
violate the so-called Seiberg bound.  They
do not have  a simple scattering interpretation.
They do play an important role, however, in providing nontrivial backgrounds,
among which the first is the two dimensional black hole [\JeYo].

The difference between the two sets of states is clearly seen in their operator
product algebras [\KlebPol]:
$$
\eqalign{
 \Psi^{(+)}_{J_1 m_1} (z)
  \Psi^{(+)}_{J_2 m_2} (w) &=  {1\over {z-w}} (J_2 m_1 - J_1 m_2)
    \Psi^{(+)}_{J_1+J_2 - 1, m_1+m_2} (w)  \cr
 \Psi^{(+)}_{J_1 m_1} (z)
   \Psi^{(-)}_{J_2,m_2} (w)
     &=  {1 \over {z-w}}
  ((J_2+1) m_1 - J_1 m_2)\Psi^{(-)}_{J_2-J_1+1,m_1 + m_2} (w), \cr
&\qquad \eqalign {J_1&<J_2+1   \cr
           |m_1+m_2|&\le J_2-J_1+1}  \cr
 \Psi^{(+)}_{J_1 m_1} (z)
    \Psi^{(-)}_{J_2 m_2} (w) &= 0,  \quad {\rm otherwise,}   \cr
 \Psi^{(-)}_{J_1 m_1} (z)
 \Psi^{(-)}_{J_2 m_2} (w) &= 0. \cr
}  \eqn \walgebra
$$

Here the positively dressed vertex operators close a $W_\infty$
 algebra while the
operator product among negatively dressed  states is trivial.

Matrix models seem to capture very well certain aspects of the theory.
In particular, the physical (positively dressed) operators appear as
$$
  Tr \left( (P+M)^{J+ m}(P-M)^{J-m}\right),  \eqn \mstatesa
$$
or
$$
  \int dx\, \psi^\dagger(x) (a^\dagger)^{J+m}a^{J-m}\psi(x) \eqn \mstatesb
$$
in terms of free fermions.
In this formulation we can very simply solve the scattering problem
to arbitrary order [\JRvT]. This represents a notable success of the matrix
model approach.

In the matrix model, however, the presence or interpretation of
the negatively dressed (singular) states is at best questionable.
For example, a naive extension $J\to -J$ in
$\mstatesb$ would lead to  singular-seeming
expressions of the type $(a^\dagger)^{-\bar n}a^{-n}$.
Apart from the mathematical difficulties involved in trying
to make sense of such operators, it is hard to see
how they
could be made to satisfy anything close to the algebra given above in
$\walgebra$.

A second basic problem seems to appear in the transition to the
(collective) field theory representation.  There the states are
represented in terms of  a massless scalar field, in particular
$$
  \int dx \int d\alpha\,
    (\alpha + x)^{J+m}(\alpha - x)^{J-m},
$$
where $\alpha_\pm = \pi_{,x}\pm \phi$.  The string coupling constant
$g_{{\rm st}} = 1/N$ now appears through the constraint on the
total number of nonrelativistic particles$\int dx\,\phi = N$. Consequently,
 the ground
state is given by the filled fermi vacuum, with excitations being particles
and holes respectively above and below the fermi surface. Perturbation theory
explores the region close to the Fermi surface. Any description
of large excitations, however, has to take into account that holes
cannot be deeper than $N$, and we expect  serious modifications
of the theory for momenta close to $1/g_{\rm st}$.  This issue is of
central relevance in connection with the evaluation of nonperturbative effects,
which in string theory are
expected to be of order $e^{-1/g_{\rm st}}$.

A proper understanding and satisfactory description of the
singular states, as well as
the nonperturbative cutoff $1/g_{\rm st}$, are therefore of
basic interest.

In what follows we shall address these questions. We shall provide a simple
physical picture in which both the singular states and the cutoff will appear
naturally. In our opinion the two issues are closely related.
In the representation that we outline a central role will be
played by quantum oscillators
with deformation parameter a root of unity.
This introduces a cutoff of  order  $1/g_{\rm st}$ and
 in addition provides a simple framework for the interpretation of the singular
(negatively dressed)
states. As we shall argue, these will  correspond to excitations
far below the fermi surface. A picture based on a q-oscillator
representation is
then offered as a simple framework in which to
study the interaction between excitations
near the Fermi surface and those which are very deep below it.

\medskip

{\chapter {From conformal field theory to free fermions}}
\medskip
The correspondence between the world sheet, conformal field theory
and the target space or matrix model description of string theory is at
present known only on a rather heuristic level. For this reason we
begin with a reexamination of what seems to be the best understood
subject: the positively dressed states.
As we have seen, analogous objects in the matrix model are given
by $\mstatesa$ or
$\mstatesb$. However, the full physical meaning of these fermionic
states has not yet been clarified.

A clue for the interpretation that follows
and much of our analysis
comes from realizing the following fact:  The continuum vertex
operator representation $\voprep$ displays a simple exponential dependence
$e^{p_\varphi \phi}$ on the Liouville mode.
This is suggestive of a hamiltonian description
in which $\varphi_0$ (the zero mode of $\varphi(z)$) plays the role of time.
In the matrix model, however,
the hamiltonian time is a priori the target space coordinate, which
would be expected to correspond  to the zero mode of the conformal
field coordinate $X(z)$.

This mismatch of space and time coordinates between the two approaches
seems to imply a serious obstacle to any comparison
between them.  However, as we shall indicate below,
in the matrix model an interchange of the roles of space and time,
at least in an analytically continued sense, at most changes the
boundary conditions in its fermionic description.
For the moment assuming this, we shall proceed by looking
for an identification between
conformal field theory states and fermionic (matrix model) excitations,
associating the zero mode of $\varphi$ to the matrix-model time coordinate
(and therefore the matrix model energy levels to
the eigenvalues $J$ of the conjugate
variable ${\hat p}_\varphi$).

Consider now in more detail the precise form in terms of oscillators
of the conformal field theory states.  For example, for the sequence
of left-moving states of the form
$$
\Psi^{(+)}_{J,J-1} \equiv
     H_-(z):e^{iJX(z)}:e^{(-1+ J)\varphi(z)}:
    \eqn \eq
$$
in the hilbert space with oscillators introduced as
$$
  \partial X (z) = \sum_n \alpha_n z^{-n-1},
$$
where
 $$
  \alpha_n = a_n \quad {\rm and} \quad \alpha_{-n} = n a^\dagger_n,
$$
  we find
$$
  \left| \Psi^{(+)}_{J,J-1} \right> = S_{2J-1}(a_n)\left|0\right>
   e^{\sqrt 2(-1+J)\varphi_0},
$$
where the $S_n$ are Schur polynomials of order $n$.

Similarly, for the
right-moving states we find
$$
  \left| \Psi^{(+)}_{J,-J+1} \right> = S_{2J-1}(-a_n)\left|0\right>
   e^{\sqrt 2(-1+J)\varphi_0}.
$$
The Schur polynomials are known to have a physical interpretation
as excitations in a free fermion theory.
In particular, $S_n(a)$, corresponding to
the young tableau with one row and $n$ columns, excites a fermion
at  $n$ steps above the fermi surface.
Similarly, the second series of states corresponds
to hole excitations.
Thus,  we see that the vertex operator representation translates into a
nonrelativistic fermion representation with the
Liouville mode playing the role of time.
As we have discussed, this fermionic picture of the vertex
operator states is dual to the standard matrix model (fermion)
picture in the sense that the roles of time and space are interchanged.

In general, it is a fermionic interpretation that brings
in the requirement of a cutoff; namely, it is
known that the sequence of polynomials that corresponds to the column states
must terminate once the column is longer than $N$;
 in other words, a hole cannot be deeper than the fermi momentum.
 This constraint  is not obviously recognizable in
conformal field theory itself.
Its implementation
will be our first concern.

Before presenting the main discussion, let us close the section
with the promised comment on how the
matrix model transforms under an interchange of space and time directions.

Taking the free fermion representation that arises in the matrix model, and
assuming periodicity in the holomorphic
representation in terms of the analytically continued $z=x+ip$, we
have, taking $H =  {1 \over 2}
 (z\partial + \partial z) = z\partial + 1/2$, the expansion
$$
  \psi = \sum_{n\in Z + 1/2} \psi_n z^{n-1/2}e^{int}.  \eqn \eq
$$

Interchanging the roles of energy and momentum
by taking $z\leftrightarrow e^{it}$, we
find
$$
\eqalign{
  \psi &\to \sum_{n\in Z+1/2} \psi_n z^{n} e^{i(n-1/2)t} \cr
     } \eqn \xtot
$$
We now have Ramond fermions up to a renaming $\psi_{n-1/2}\to\psi_n$ of the
modes.
Thus, we see that in the free fermion description of the matrix model,
interchanging the time and space directions as above will take
Neveu-Schartz (periodic) fermions into Ramond
(antiperiodic) fermions (this is also
consistent with the fact that the plane to
cylinder mapping ($z\to e^{it}$) changes the
periodicity of the fermions [\Ginsparg]).

We remark that, by extension of the above argument,
introducing a nonzero chemical potential $\mu$
in the matrix model Hamiltonian would translate in the
rotated system into having twisted
fermions with the phase of the twist depending on $\mu$.  In this paper
the dependence on the number of
fermions $N$ will be taken into account explicitly through
the use of q-oscillators and we will not need a nonzero $\mu$.  Allowing
$\mu$ to vary would indeed correspond to allowing an additional degree of
freedom, over and above those that are a priori provided by the matrix
model.

\medskip
{\chapter {Quantum oscillators and the cutoff}}
\medskip

As we have seen, the fermionic interpretation of the conformal field
theory states  very naturally leads to the suggestion that there should
be an explicit  cutoff present on the space of  states themselves,
a constraint which is not
a priori visible in the conformal field theory
itself in its usual formulation. In particular, the holes should
not be deeper than $N$ (the total number of fermions, or order of the
matrix), which is  $1/g_{\rm st}$ in terms of the string coupling.
If we want left- and right-moving
states to appear symmetrically in the theory,
we are motivated to impose the same cutoff on the momenta of the
particle states.

In order to ensure that there arise no excitations deeper than $N$,
we are led to suggest a representation in terms of q-oscillators,
which we motivate as follows:

Consider a hole in a first quantized formalism, and
denote the corresponding creation and annihilation operators
by $a^\dagger$ and $a$.
The fact that a hole cannot be excited deeper than $N$ will be imposed
by requiring
$$
(a^\dagger)^N = a^N = 0.
$$
This imposes the need to modify the standard commutation relations.
One obvious candidate would be
$$
 [a, a^\dagger] = 1 - (\hat N +1) \delta_{\hat N, N-1},
   \eqn \acomm
$$
where $\hat N = a^\dagger a$ is the number operator.  The algebra of
operators becomes finite-dimensional,  with a basis given by the $N^2$
generators $(a^\dagger)^m a^n$, $m,n = 0,...,N-1$.  However, due to the
second term on the right hand side in $\acomm$,
it is not easy to see how to define a
consistent coproduct [\Zha], which will be needed to extend the algebra to a
multi-particle representation in the second quantized theory.
However, we can
already see possible candidates for the singular states as being given
by the ones near the top of the ``stack",
namely $(a^\dagger)^{N-n} a^{N-m}$.

One solution to the problem of the coproduct is given
in terms of q-deformed oscillators satisfying
$$
  \eqalign {
  bb^\dagger - q^2 b^\dagger b &= 1,\cr
  [\hat N, b^\dagger] = [b, \hat N] &= 1,  \cr
} \eqn\quantumb
$$
where $q$ is a root of unity, satifying $q^{2N} = 1$, $\hat N$ is the number
operator and $b\left|0\right> = \hat N \left|0\right> = 0$.
 It can be shown that in this
representation we have $b^N = (b^\dagger)^N = 0$, which incorporates
the cutoff.  It is easy to convince ourselves that we can express such
oscillators in terms of the $a$'s of $\acomm$ above by defining, for example,
$$
\eqalign
{
  b^\dagger = \sqrt { {[\hat N]}\over {\hat N}} a^\dagger,    &\qquad
  b = a \sqrt {{[\hat N]}\over {\hat N}},     \cr
  \hat N &\to \hat N, \cr
  [p] &\equiv  {{q^{2p-1}} \over {q^2 - 1}}. \cr
}
$$
For $q$ a phase, as it will be in our case,
$b^\dagger$ will no
longer be the hermitian conjugate of $b$.  A related set of
oscillators with simpler hermiticity properties is given by
 $a= q^{-\hat N/2} b$,
$a^\dagger = b^\dagger q^{-\hat N/2}$, where now $(a^\dagger)^\dagger = a$ and
we get the q-commutation relations
$$
\eqalign{
  a a^\dagger - q a^\dagger a &= q^{-\hat N},  \cr
  a a^\dagger - q^{-1}a^\dagger a &= q^{\hat N}.  \cr
}  \eqn \quantuma
$$
For a review, see
[\BaFlo,\McFBP].  Discussion of the coproduct will be postponed
to section 4.

\noindent
Define as a basis for the operators on the one-hole hilbert space
$$
  O_{Jm}  \equiv (b^\dagger)^{J+m} b^{J-m}.   \eqn \ob
$$
These operators satisfy a deformed version
of a W-algebra:
Following Zha [\Zha], we  define a
deformed  version of the commutator, which we shall call  a q-bracket, as
$$
\eqalign{
  [O_{Jm}, O_{J'm'}]_q &\equiv q^{-2 (J'm-Jm')}\, O_{Jm}\,O_{J'm'}
           - q^{2 (J'm-Jm')} \,O_{J'm'}\,O_{Jm}   \cr
     &= ([J' + m']_q \,[J-m]_q \,q^{-2m} - [J+m]_q \,[J'-m']_q \,q^{-2m'})
          \,O_{J+J'-1, m+m'},  \cr
}  \eqn \QuantumW
$$
up to terms proportional to $O_{J+J'-s}$ with $s\ge 2$, and
where
$$[x]_q \equiv { {q^N - q^{-N}} \over {q - q^{-1}}}.$$
The motivation for the powers of $q$ appearing in the
definition of the q-bracket is that they are
what is needed for the term proportional to
$O_{J+J',m+m'}$ to vanish on the right hand side.
For the coefficients of all the lower order terms we refer the
reader to [\Zha].

The undeformed W-algebra is obtained as a limit of the
q-deformed version above as $q\to 1$.  In particular,
in our case $q= e^{\pi i /N}$, where $N$ is the cutoff, so that
when $N$ becomes
large and $x<<N$, we get $[x]_q \to x$, while the q-bracket reduces to
the ordinary commutator,
 so that for operators with
sufficiently low $J$-values
$$
\eqalign{
  [O_{Jm}, O_{J'm'}]
     &\to ((J' + m') (J-m) - (J+m) (J'-m'))
          O_{J+J'-1, m+m'}  \cr
 &= 2 (m'J-mJ')
          O_{J+J'-1, m+m'}  \cr
    }
$$

As a bonus, in this picture we now get candidates for the negatively
dressed (singular) states with the correct commutation relations
$\walgebra$ in the limit.  Indeed, defining
$$
 \bar O_{Jm} \equiv O_{N-J-1, m},\qquad |m|\le J
$$
we find, using $q^{2N} = 1$, that for large $N$ the q-bracket
between $O$'s and $\bar O$'s, and indeed between two $\bar O$'s,
reduces to the ordinary commutator.  Furthermore, from
$[x]_q = [N-x]_q \to x$, it follows that
$$
\eqalign{
  [O_{Jm}, \bar O_{J'm'}]_q
     &= ([N - J'-1+ m']_q [J-m]_q    \cr
        &\qquad - [J+m]_q [N -J'-1-m']_q)
          \bar O_{-J+J'+1, m+m'}  \cr
 &\to - 2 ((J'+1)m + Jm')
          \bar O_{-J+J'+1, m+m'},  \cr
  } \eqn \one
$$
as long as $J<J'+1$ and $|m+m'|\le J'-J+1$.  If not, the result will be
$0$, because in the  algebra $\ob$ there are no operators
with power of $b$ or of $b^\dagger$ larger than $N-1$.  By the
same token,
$$
  [\bar O_{Jm},\bar O_{J'm'}] = 0.
\eqn \two
$$
With this we have demonstated that the algebraic structure of 2-d string
theory vertex operators is completely reproduced in the q-oscillator
phase space.  This is particularly nontrivial for the singular
(negatively dressed) states.

The algebra $\one$ and $\two$ is the same  as the
one $\walgebra$ obtained in the conformal field theory approach.
However, so far it only contains the hole contribution, and in the full
theory we have to include the particles.

To do this, we introduce another set of q-creation and annihilation
operators ${\bar b}^\dagger$ and $\bar b$ associated to the particles.
In the following, we will use the notation
  $$
\eqalign {
    \partial, z &\equiv b, b^\dagger,  \cr
    \bar \partial, \bar z &\equiv \bar b, {\bar b}^\dagger  \cr
}  \eqn \zb
$$
suggestive of a q-deformed holomorphic-antiholomorphic representation.
Following [\BaFlo], we extend the commutation relations to
the $z$, $\bar z$ system by requiring associativity
of the differentiation rules such that the following braiding relations
are satisfied:
$$
 \eqalign{
   &z\bar z = q^2 \bar z z  \cr
   &\partial \bar\partial = q^2 \bar\partial \partial,\quad
       \partial \bar z = q^{-2} \bar z \partial,  \quad
       \bar \partial z = q^2 z \bar \partial \cr
   &\partial z = 1 + q^{-2 } z\partial, \quad  \bar \partial \bar z = 1 + q^2
\bar z
       \bar \partial. \cr
} \eqn \zbraid
$$

To generalize the expression for the W-charge $\ob$ to
two oscillators, one might try the expression
$ (\partial+{1 \over 2} \bar z)^{J+m} (\bar\partial  - {1 \over 2} z)^{J-m}$.
This works
in the undeformed case $q=1$.  However, in the deformed case
 $ (\partial + {1 \over 2} \bar z) (\bar\partial - {1 \over 2} z) -
q^2 ({\partial
\bar z - {1 \over 2} z)} (\partial z + {1 \over 2} \bar z) \ne 1$.  This
happens because $\partial$ and $\bar\partial$ pick up the ``wrong" power of
$q$ when the are interchanged, and thus
the terms in $\partial\bar\partial$ do not cancel.  This problem
can be avoided by taking
as generators the expressions $(\bar \partial + {1 \over 2} z)$ and
$\bar z$, which do satisfy the  q-commutation relation
$\quantumb$.  We are
therefore motivated to consider the following candidate
for a particle-hole
q-deformed W-algebra:
$$
  O_{Jm} = (\bar \partial + {1 \over 2} z)^{J+m} {\bar z}^{J-m}.  \eqn\nonsymm
$$
These generators indeed satisfy the above q-deformed
W-algebra $\QuantumW$.

The representation $\nonsymm$ suggests the following correspondence
with the conformal field theory states as discussed in paragraph 2:
to leading order in $z$ and $\bar z$ we have
$$
  O_{Jm}  \approx z^{J+m} {\bar z}^{J-m},  \eqn \ozz
$$
which can be viewed as representing a particle-hole pair at distances
$J+m$ and $J-m$ respectively above and below the fermi surface.  The
total energy
of this state is given by $(J+m) - 1/2 + (J- m) -1/2 = 2J-1$,
while the momentum is
$J+m - (J-m) = 2m$.  This is  consistent with our picture
of taking the liouville direction as time.

In this representation, we can now interpret
the candidate for the black hole
operator, namely
$$
  \bar O_{0, 0} \approx O_{N-1, 0} = (\bar \partial + {1 \over 2} z)^{N-1}
                      {\bar z}^{N-1} \approx z^{N-1}{\bar z}^{N-1},
$$
as the state with maximum possible energy $2J-1$ that can be excited in
the system, describing a hole at depth $N$ below
the fermi surface, combined with a particle at the distance $N$ above the
fermi surface.

It is instructive to compare our singular
states to representation given by Witten [\Witten].
A connection is seen as follows:  A natural definition
of integration over a $q$-commuting variable $z$ satisfying
$z^N = 1$ is given by [\BaFlo]
$
  \int dz \, z^{N-1} = 1
$.
On this space the delta function is therefore given by $\delta (z) = z^{N-1}$.
Now, the negatively dressed states are, from $\ozz$
$$
\eqalign{
     \bar O_{Jm} &= O_{N-1-J,m}   \cr
                 &\approx z^{N-1-J+m} \bar z^{N-1-J-m}  \cr
                 &\approx \partial^{J-m} \bar \partial^{J+m}
                       \delta (z, \bar z), \cr
}   \eqno \eq
$$
which gives an expression similar to that obtained by Witten.

In summary, we have shown in this section how the phase space of a
q-oscillator accomodates both the positively dressed and the negatively
dressed states in two-dimensional string theory.  We have shown that the
algebraic structure, i.e., their commutators, is precisely reproduced,
with the positive states closing a $W_\infty$ algebra while the commutators
of negative states are seen to vanish.  Consequently, this
q-oscillator picture provides a framework in which the interaction of
all these degrees of freedom can be studied.  We emphasize that the
negatively dressed states were accomodated not through adding extra
degrees of freedom but based on our interpretation that they correspond
to states deep below the fermi surface.  The use of q-oscillators
allowed us to exhibit these states explicitly.

\medskip
\chapter{Second-quantized representation}
\medskip

As a first step towards understanding the theory at the second-quantized
level, we shall now
exhibit a second quantized version in terms of free fermions
of the algebra $\QuantumW$ associated to the hole states only.
We shall leave to future work
the discussion of combining particles and holes in the full theory, i.e.,
second quantizing  $\nonsymm$.

In this spirit, we are therefore motivated to search for a second quantized
 q-deformed W-algebra in terms of $N$ fermi modes and their
conjugates, representing hole creation and annihilation operators.
In the following, we will use the q-calculus as described in
[\BaFlo].

Let $z$ and $\partial$ be $q$-holomorphic coordinates as defined
in $\zb$,
with $q^{2N} = 1$.
Take $N$ femionic modes $\psi_{-n}$, $n=0,...,N-1$,
 and their conjugates,
satisfying the usual anticommutation relations $\{\psi_{-m}, \psi^\dagger_n
\} = \delta_{m+n}$.  We now define q-holomorphic fields
by the expansions
$$
  \eqalign{
   \psi (z) &= \sum_{n=0}^{N-1}\psi_{-n} z^n  \cr
   \psi^\dagger (z) &= \sum_{n=0}^{N-1} z^{N-1-n} \psi^\dagger_n, \cr
}   \eqn \psis
$$
Here the modes of $\psi$ and $\psi^\dagger$ are taken to commute with
$z$.
The reason for the peculiar power of $z$ in the expression for $\psi^\dagger$
comes from the definition of the $dz$-integral as $\int dz\, z^{N-1} =1$
(see [\BaFlo]) so that $z^{N-1}$ plays a role similar
to that of $z^{-1}$ in the commuting theory.   We will discuss the
generators in terms the q-position space representation below, but first
let us try to guess their form in the momentum representation.

Now, in a second-quantized representation of the q-deformed
W-algebra, the coproduct
becomes important, as it allows us to consistently define the action
of the generators on multi-particle states without spoiling the
q-commutation relations.   In our case, we can check that the following
coproduct preserves the q-commutators:
$$
  \Delta ( (z^{n+k}\partial^k) = z^{n+k}\partial^k \otimes q^{2k\hat N}
     + q^{2k\hat N} \otimes z^{n+k}\partial^k.
$$
This coproduct is different from the one described in [\Zha].  However
because of its symmetry, it can now easily be extended to multi-particle
states, motivating the following expression for the second-quantized
generator:
$$
   W^{(k)}_n = \sum_p \psi^\dagger_{p+n} \,[p]\,
            [p-1] ... [p-k+1]\, q^{2kN_\psi}
                 \, \psi_{-p},    \eqn \second
$$
where $N_\psi = \sum n\, \psi^\dagger_n \,\psi_{-n}$, and where the ranges
of summation are taken such that $p$ and $p+n$ fall inside $[0,...,N-1]$.
Here we remind the reader that $[x] \equiv (q^{2x}-1)/(q^2-1)$.

To see that this works, it is straightforward to check the
crucial fact that, because of
the presence of the $q^{2kN_\psi}$, the quartic terms in the fermion
oscillators cancel in the q-bracket (adjusted here for the notation in
terms of $k$ and $n$)
$$
  q^{(nk'-n'k)}\,W^{(k)}_n W^{(k')}_{n'}
  - q^{-(nk'-n'k)}\,W^{(k')}_{n'} W^{(k)}_n,
$$
which therefore reduces to
$$
\eqalign{
 \sum_p \psi^\dagger_{p+n+n'} \,(
     & p^{(nk'-n'k)} \,[p+n'] ... [p+n'-k+1]\,[p] ... [p-k'+1]
  \cr
     - & p^{-(nk'-n'k)}\, [p+n] ... [p+n-k'+1]\,[p] ... [p-k+1]
   )\,  q^{2(k+k')N_\psi} \,\psi_{-p}. \cr
}
$$
Now the term in the inner brackets is just
$ [z^{n+k}\partial^k, z^{n'+k'}\partial^{k'}]_q$ applied to the basis element
$z^p$,
so that the correspondence with the first-quantized algebra follows.

It is important to realize that the generators $\second$ of the
second-quantized q-W-algebra are expressed in terms of ordinary fermionic
operators with no mention of any exotically commuting quantities.
However, these generators satisfy a simple algebra with respect to q-brackets
instead of ordinary commutators.
Even so, we stress that the above expressions do not contain any degrees
of freedom other than those already present in the ordinary fermion
theory.

It is possible to formulate the theory also in q-position space in
a very nice form.   Taking the fields as a function of $z$ as in $\psis$,
we define
$$
  W^{(k)}_n \equiv \int dz\, \psi^\dagger(z) z^{(n+k)}\partial^k\psi,
$$
with $z$ and $\partial$ as in $\zb$.
This leads to the expression $\second$ if we define the $z$-integral
as $\int dz z^{N-1} = 1$ (see Floratos), and if in addition we take
the following braiding relation between $\psi$ and $\partial$:
$$
 \partial \psi_{-n} = q^{-2n} \psi_{-n} \partial,   \eqn \psizbraiding
$$
(which can always be compensated by the redefinition
$
  \partial \to q^{2N_\psi}\partial
$).

Now, to calculate the q-bracket of two such expressions in $z$-space,
take two ``time slices" indexed by $z$ and $w$ and following
[\ChaiDem] impose the following braiding
relations:
$$
  zw = wz, \quad z\partial_w = q^{-2}\partial_w z, \quad w \partial_z =
q^{-2}\partial_z w,
   \quad w\,dz = dz\,w, \quad \partial_w dz = q^2 dz\,\partial_w, ...
  \eqn\zw
$$
which will give us consistent associative
differentiation and integration rules.  Note that these braiding relations
are different from the ones between $z$ and $\bar z$ in
$\zbraid$ (which formally
can be considered as conjugate variables on the same ``time slice").
Now, being careful with powers of $q$ picked up when commuting
integrations, etc., a straightforward but tedious calculation gives
indeed
$$
\eqalign{
  [W^{(k)}_n, W^{(k')}_{n'}]_q =
   \int dz\, \psi^\dagger (z) \left( [z^{n+k}\partial^k,
      z^{n'+k'}\partial^{k'}k]\right) \psi(z),
}
$$
in other words, the quartic terms cancel due to the
braiding relations $\psizbraiding$ and $\zw$ and we are left with an
isomorphism between the field theoretic and the $z$, $\partial_z$
representation.
The proof of the above proceeds analogous to the
commuting case by noting that according to the
$z$-integration we have
$$
  \delta (z-w) = \sum_{p=0}^{N-1} z^p w^{N-1-p} = \{\psi (z), \psi^\dagger
     (w)\}.
$$

We close with a remark on dynamics:
One might  decide to take the above
formulation of the theory for finite $N$ as fermions defined on
a quantum holomorphic space as fundamental, and define the
dynamics accordingly.  For related discussions of the
natural appearance of quantum groups in string theory, see [\BabGerAGS].

In this spirit, let us indicate in broad
terms how the calculation of in-out reflection coefficients
for the fermions may
be approached in such a theory.
One way to derive these coefficients is, following
[\OLough], to write the fermions in a
holomorphic basis in terms of powers of
$z = x+ip$ (for simplicity we now only consider
the right side up harmonic oscillator), in which case the in-out
transformation is obtained by reexpressing the in-field in terms of the
canonically conjugate coordinates $\bar z = x-ip$.  The in-out
bogoliubov transformation thus becomes a fourier transform.

We have described a formulation in terms of fermions on a noncommuting
q-holomorphic space.  If q-space is fundamental, we
would expect the reflection coefficients to be given by a q-fourier
transform.  A possible definition of such a transform is given
by
$$
  F(z) \equiv \int d\bar z\, e^{z\bar z}_q f(\bar z),
$$
where the q-exponential is defined as [\BaFlo]
$$
 e_q^{az\bar z} \equiv \sum_{p=0}^{2N-1} a^p  {{(\bar z z)^p}\over{[p]!}},
$$
where one can check that the inverse transform is given by
$$
  f(\bar z) = \int dz \equiv e_{q^{-1}}^{-q^2\bar z z} F(z),
$$
which is reasonable if one takes into account that $q\to q^{-1}$ if
we interchange $z$ with $\bar \partial$ and $\bar z$ with $\partial $ in the
q-commutation relations.

Now a careful evaluation of the q-fourier transform of
a field $\psi_{-k}{\bar z}^k$ commuting with $z$ and $\bar z$ gives
$$
  \int d\bar z\, e^{-q^2 z\bar z}_{q^{-1}}\, {\bar z}^k\, \psi_{-k}
   = {1 \over{[N-k-1]!}} z^{2N-1-k} \psi_{-k},
$$
while for its conjugate $\psi^\dagger_{k}\,{\bar z}^{2N - 1 - k}$
we should define it as
$$
 \psi^\dagger_k\,
 {\bar z}^{N-1-k} \,e_q^{\bar z z}  \, d\bar z \int_{\leftarrow}
   = {q^{k(k+1)} \over {[k]!}} \,\psi^\dagger_k\, z^k,
$$
where the right to left integral is defined in the obvious way.
But we can show that ${q^{k(k+1)} / {[k]!}} = [2N-1 - k]!$ up to
a $k$-independent factor, in other words, $\psi$ and $\psi^\dagger$
transform dually.  This is needed for one obvious consistency
condition, namely that
the operator $N_\psi = \sum n \psi^\dagger_n \psi_{-n}$ be invariant
under an in-out transformation, to be satisfied.

The q-reflection coefficients are seen to be simple generalizations
of what one would obtain in the harmonic oscillator.  This analysis
would correspond in the $q\to 1$ limit to the case of the
right side up harmonic oscillator potential.  It would be interesting
to carefully study the non-analytically-continued
case.  These and related questions are left to future work.

\medskip
\chapter {Acknowledgements}
\medskip
One of us [AvT] would like to thank the Physics Department of Brown
University for their kind hospitality during my stay, as well as the
Institute of Physics of the University of Hannover, Germany, under
whose auspices some of the initial work for this article was done.

\refout

\end